\documentclass{aa}

\usepackage{graphics,times}

\begin{document}

\thesaurus{1(03.20.2; 11.06.2; 11.14.1; 11.16.1; 11.19.1; 11.09.1 NGC 1068)}

\title{Diffraction-limited 76\,mas speckle masking observations of the core of 
NGC 1068 with the SAO 6\,m telescope
\thanks{Based on data collected at the Special Astrophysical Observatory, 
Russia}}

\author{M.\ Wittkowski\inst{1,2} \and Y.\ Balega\inst{3} \and T.\ 
Beckert\inst{2} \and W.J.\ Duschl\inst{2,1} \and K.-H.\ Hofmann\inst{1} 
\and G.\ Weigelt\inst{1}}

\offprints{Markus Wittkowski, Bonn; E-Mail: mw@speckle.mpifr-bonn.mpg.de}

\institute{Max-Planck-Institut f\"ur Radioastronomie, Auf dem H\"ugel 69, 
D-53121 Bonn, Germany
\and
Institut f\"ur Theoretische Astrophysik, Tiergartenstra{\ss}e 15,
D-69121 Heidelberg, Germany
\and
Special Astrophysical Observatory, Nizhnij Arkhyz, Zelenchuk region, 
Karachai-Cherkesia, 357147, Russia}

\date{Received \dots ; accepted \dots}

\maketitle

\markboth{M.\ Wittkowski et al.: Speckle Masking observations of the core of 
NGC 1068}{M.\ Wittkowski et al.: Speckle Masking observations of 
the core of NGC 1068} 

\begin{abstract}
We present the first K-band bispectrum speckle interferometry of NGC\,1068 
with an angular resolution of $76\,{\rm mas}$ ($\sim 5.5$\,pc). This
angular resolution allows us to attribute the measured flux
to only one of the nuclear sources seen at radio wavelengths. 
The observed decreasing visibility function suggests that the dominant 
central core is probably not an unresolved point source, but slightly resolved 
with a FWHM diameter of $\sim 30$\,mas $\sim 2$\,pc for an assumed Gaussian 
intensity distribution. This 30\,mas object is possibly the nuclear torus 
and/or a scattering halo.

We discuss different contributions to the observed K band flux.
Between 5\,GHz and the K-band the spectrum of this component is close to a 
$\nu^{1/3}$ proportionality. 
In addition to the standard interpretation of a hot dust torus surrounding
the nucleus of NGC\,1068, one cannot exclude the possibility that a 
sizeable fraction of the nuclear flux reaches us via a scattering halo.
This then would allow us to determine physical parameters
of the nuclear source.

\end{abstract}


\section{Introduction}
The Seyfert 2 galaxy NGC\,1068 is one of the nearest and brightest Seyfert 
galaxies and thus one of the closest candidates for an actively accreting
supermassive black hole. Its distance is about 15\,Mpc, corresponding to 
73\,pc/$^{\prime\prime}$.
Cores of Seyfert galaxies are classified as types 1 and 2, with type 1 
exhibiting broad and narrow lines while the spectra of type 2 show only 
narrow lines. Unified theories of AGN propose that all AGN harbour a 
continuum source surrounded by a dusty molecular torus. Depending on the 
observer's viewing angle this torus either obscures the view on 
the inner source
(type 2), or it does not (type 1) (Antonucci and Miller 1985). 
For NGC\,1068 Bailey et al. (1988) find from NIR spectroscopy a 
rather low ${\rm A_V}$,
perhaps as small as ${\rm A_V} \sim 10^{\rm m}$.
The extinction is 
wavelength-dependent and much less in the near infrared, allowing for 
a deeper look towards type 2 cores in the
IR. Accordingly, the first high-resolution IR observations 
of NGC 1068 exhibited a spectacular compact central IR core and an
underlying galaxy (Chelli et al.\ 1987; Blietz et al.\ 1994; 
Tacconi et al.\ 1994; Young et al.\ 1996; 
Weinberger, Neugebauer, Matthews \ 1996; 
Quirrenbach, Eckart, Thatte \ 1997; Thatte et al.\ 1997).

Recent investigations of the center of {\it our\/} Galaxy indicate that the 
difference between it and a Seyfert core is not of a generic nature but rather 
a question of the {\it current\/} level of activity (Mezger, Duschl, Zylka \ 1996 =
MDZ96). For our Galactic Center it was shown by Duschl and Lesch (1994) and 
-- in more detail -- by Beckert et al.\ (1996 = BDM96) that the radio-IR 
spectrum of the central source Sgr A* can be explained as optically thin 
synchrotron radiation of quasi-monoenergetic relativistic electrons. The 
spectrum is characterized by a flux density $F_\nu \propto \nu^{1/3}$ between 
a maximum flux in the FIR range and a low frequency turnover at $\sim 1$ GHz 
due to synchrotron self-absorption (SSA).

Recently, Muxlow et al.\ (1996 = MPH96) have presented spectacular MERLIN 
interferometry
observations of the central radio structure of NGC\,1068 between 5 and 
22\,GHz. They have separated the core into 5 components with a minimum 
distance between individual components of $\sim 100$ mas. All but one of 
the components show negative spectral indeces ${\rm d} F_\nu / {\rm d} 
\nu$ between $-0.33$ and $-0.88$, while the remaining component shows a 
positive index of $+0.31$. MPH96 identify this component as the true 
center of NGC\,1068, in so far as resembling the nuclear source of our Galaxy,
Sgr A*.

In this letter we present speckle masking bispectrum observations of the 
core of NGC\,1068
in the K-band with a resolution of 76 mas, give the flux at
2.2 $\mu$m, and discuss the nuclear radio-IR spectrum of this Seyfert
galaxy.

\begin{figure}
\centerline{\resizebox{0.8\hsize}{!}{\includegraphics{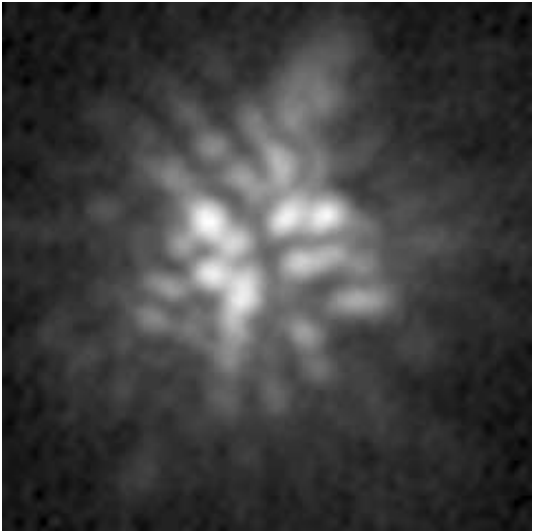}}}
\caption{One of our 778 speckle interferograms of NGC\,1068 taken through
a K-band filter, demonstrating that it is possible to obtain speckles 
with high signal-to-noise ratio from a K $\sim 8$ source, even under not 
very good seeing conditions. The observations were 
made using a NICMOS\,3 array with 200 ms exposure time per frame and multiple
read-out (4$\times$). Furthermore, the high-contrast speckles show that the 
NGC\,1068 core is very compact since the speckles look like speckles of a
point source. The shown field of view is 1\farcs 85 $\times$ 1\farcs 85. 
\label{si} }
\end{figure}

\begin{figure}
\centerline{\resizebox{0.9\hsize}{!}{\includegraphics{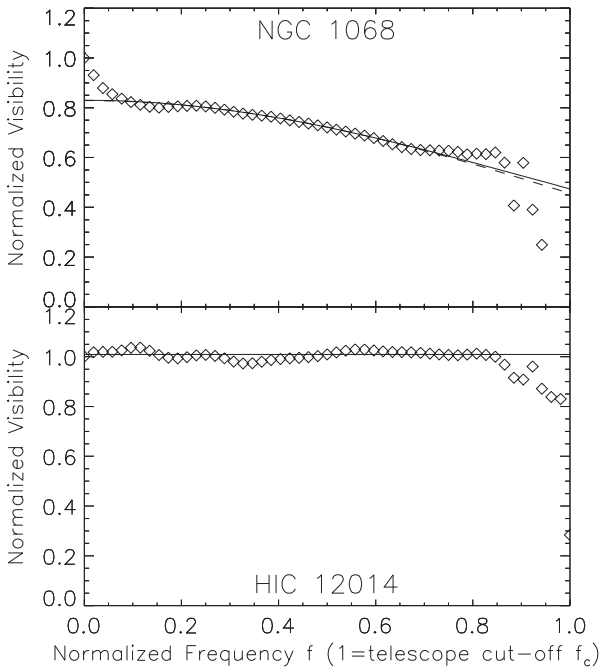}}}
\caption{Azimuthally averaged visibilities of NGC\,1068 (top) and
the unresolved reference star HIC\,12014 (bottom). The diamonds indicate
the observed visibilities, the solid lines the Gaussian fits, while the 
dashed line is a UD fit. The fit range is between 20\,\% and 70\,\%
of the telescope cut-off frequency $f_{\rm c}$. The visibility errors 
at 20\,\%, 50\,\%, 70\,\%, 80\,\% and 90\,\% of $f_{\rm c}$ are 
$\pm$ 0.05, $\pm$ 0.05,$\pm$ 0.08, $\pm$ 0.15, $\pm$ 0.20, respectively. 
\label{vis}  }
\end{figure}

\section{Observations and data reduction}
The NGC\,1068 speckle interferograms were obtained with the 6 m telescope 
at the 
Special Astrophysical Observatory (SAO) in Russia on October 3, 1996. 
The speckle interferograms were recorded with a 256$\times$256 pixel 
NICMOS\,3 camera through a standard K-band filter with center 
wavelength of 2191 nm and FWHM bandwidth of 411 nm. A typical speckle frame is 
shown in Fig. \ref{si}. The exposure time per frame was 200\,ms,
K-band seeing was about 1\farcs 5. 778 object speckle interferograms and 526 
reference star (HIC\,12014) interferograms were recorded. The scale was 
30.82 mas/pixel and the field of 
view 7\farcs 9 $\times$ 7\farcs 9. Diffraction-limited images were 
reconstructed
from the speckle data using the speckle masking method (Weigelt 1977;
Lohmann, Weigelt, Wirnitzer\ 1983; Pehlemann, Hofmann, Weigelt\ 1992). The process
includes the calculation of the average power spectrum and the average
bispectrum and the subtraction of the detector noise terms from those.
The 4-dimensional bispectrum of each frame
consisted of $\sim$ 49 million elements. No postprocessing by image 
restoration methods was applied to the speckle masking reconstructions.
For the calibration of the flux the photometric standard star HIC\,110609
($=$ BS 8541), chosen from Elias et al. (1982), was observed on 
September 30, 1996. We adopt the HIC\,110609 flux to be 12.57 Jy at 2.2 $\mu$m.

\section{Results}
The top diagram of Fig. \ref{vis} shows the azimuthally 
averaged visibility function of NGC\,1068 together with the corresponding 
uniform-disk
(UD) and Gaussian
fits. The object visibility function was obtained by compensating both, 
the seeing-calibrated
speckle transfer function and an additional transfer function due to the
different spectra of NGC\,1068 and the reference star HIC\,12014 (K0).
On the bottom diagram the analogous plots are shown for the reference
star HIC\,12014. The HIC\,12014 visibility function was obtained by reducing 
two different data sets of HIC\,12014. 
The reconstructed visibility function of NGC 1068 decreases to about
50\,\% of the zero-frequency value at the diffraction cut-off 
frequency $f_{\rm c}$.
The low-frequency peak is caused by the underlying galaxy. It is not yet
clear whether the short horizontal part of the visibility function at
$f>0.7\,f_{\rm c}$ is caused by an additional point 
source or by an overestimation of the transfer function due to the different
spectra of NGC\,1068 and the reference star. The decrease
of the NGC\,1068 visibility function is much larger than the 
errors (see caption of Fig. \ref{vis}). For example,
between 10\,\% and 70\,\% of $f_{\rm c}$ the visibility function
decreases by about 25\,\%, which is approximately three times larger than 
the 8\,\% error at 70\,\% of $f_{\rm c}$. Nevertheless, future
observations will be useful to confirm the result and to improve 
the accuracy.
  
The diameter of the resolved NGC\,1068 core was derived from the 
reconstructed visibility function by fitting symmetric 
uniform disks (UD) and Gaussian models. 
The fits were carried out in the range between 20\,\% and 70\,\% of
$f_{\rm c}$.  
The fitted diameters $d_{\rm Gauss}$ (FWHM)  and $d_{\rm UD}$ 
of the Gaussian and the UD model are $d_{\rm Gauss}=  
30\,{\rm mas}\pm 8\,{\rm mas}$ and $d_{\rm UD} = 50\,{\rm mas}\pm 
15\,{\rm mas}$. These diameters correspond to 2.2\, pc $\pm$ 0.5\, 
pc and 3.7\, pc $\pm$ 0.8\, pc, respectively. 
An alternative model interpretation of the decreasing visibility 
function is, for example, an object
slightly larger than the above diameters plus an unresolved central object or 
an object which is unresolved in one direction ($\sim$ EW), but resolved in the
direction perpendicular to it ($\sim$ NS).

\begin{figure}[tb]
\centerline{\resizebox{0.8\hsize}{!}{\includegraphics{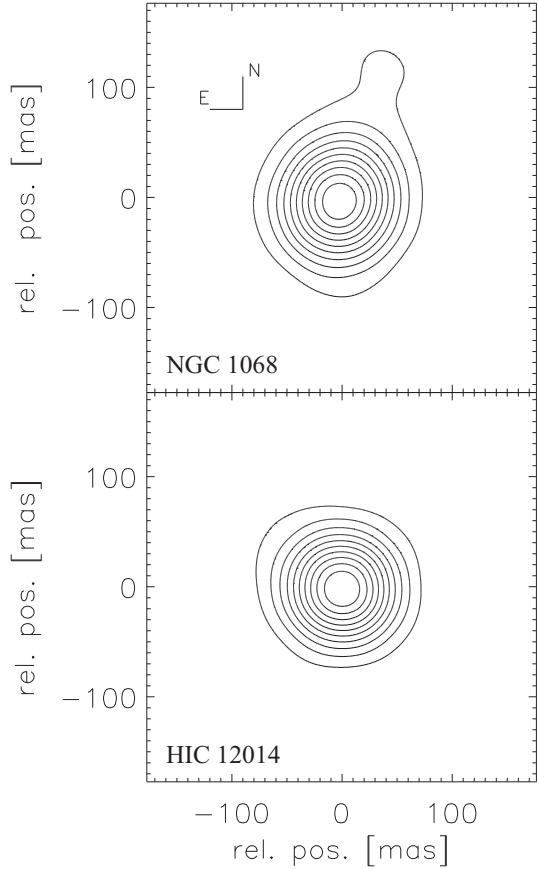}}}
\caption{Diffraction-limited speckle masking reconstruction of  NGC\,1068 (top)
and the unresolved star HIC\,12014 (bottom). The contours are plotted from
6\% to 100\% of peak intensity in 10 steps. \label{rek}}
\end{figure}

\begin{figure}
\centerline{\resizebox{0.9\hsize}{!}{\includegraphics{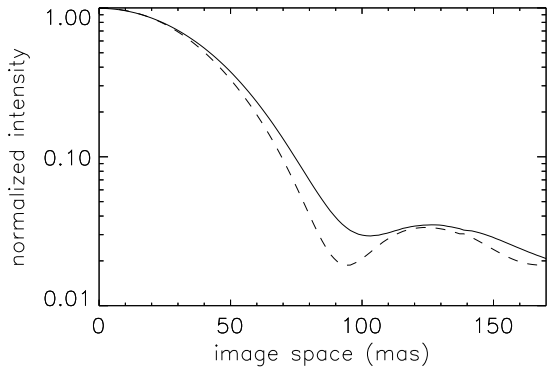}}}
%\vspace{-3mm}
\caption{Azimuthally averaged radial plots of the reconstructed images of 
NGC\,1068 (solid line) and the reference star HIC\,12014 (dashed line). 
\label{rad}}
\end{figure}

Figure \ref{rek} shows the diffraction-limited speckle masking reconstruction
of NGC\,1068. The resolution of the reconstructed image 
is 76\,mas. It shows an elongated structure in the northern direction, 
i.e. approximately the position of the radio jet.  
Figure \ref{rad} shows the azimuthally averaged radial plots of the 
reconstructions.

Photometry was performed by comparing the integral intensities of the
long-exposure images of NGC\,1068 and the photometric standard 
star HIC\,110609.
It yields for the flux $F_{\nu}$ of NGC\,1068  
in the K-band the value
$F_{\rm K} = 650\,{\rm mJy} \pm 200\,{\rm mJy}$. 
>From this value we have to subtract the flux
of the underlying faint extended component in order to get the flux 
$F_{\rm K}^{\rm 30\,mas}$ of only 
the 30\,mas component. We determine the contribution of this 
extended component from the zero-frequency visibility peak discussed
above to 20\% $\pm$ 10\% of $F_{\rm K}$ . This results in 
a flux from the 30\,mas component of $F_{\rm K}^{\rm 30\,mas} = 
520\,{\rm mJy} \pm 210 \,{\rm mJy}$.

\section{The radio--IR spectrum of the inner parsecs of NGC\,1068}

MPH96 showed that the central $\sim 2^{\prime\prime}$
have substructures on a scale of 100\,mas requiring observations with 
a resolution of better than $100 \,{\rm mas}$ to separate the true nuclear 
spectrum from that of surrounding sources. Only a few
published radio flux determinations of the nucleus of NGC\,1068 
have a sufficiently high angular resolution to allow the separation of 
individual nuclear components and thus to use it for physical 
investigations of the nucleus' property.
Fortunately, our speckle observations have the required high angular
resolution (76 mas). Therefore, our resolution would allow the 
separation of the
individual core components discussed by MPH96 if present in the
IR. We assume that the single source that we have observed in the K-band is 
the same as the true nucleus observed by MPH96. Our observations
constitute an upper limit to the volume from which the above
determined $520\,{\rm mJy} \pm 210 \,{\rm mJy}$ originate. 

Usually, the K band flux from nuclei of Seyfert\,2 galaxies is attributed
either to a warm dust torus or a compact nuclear stellar cluster or a
combination of the two (e.g., Thatte et al.\ 1997). 
If our source is the torus that is held 
responsible for the different appearences of Seyfert\,1 and 2 galaxies, our 
observation constitute the first determination of a torus size. 
To clarify the nature of the radiating source, further spectroscopic 
and polarimetric measurements with a similarly high angular resolution are 
necessary.

However, a combination of the flux measurements and nuclear source 
identification by MPH96 in the
radio frequency regime with our observation makes it intriguing to speculate
whether a sizable 
fraction of $F_{\rm K}^{\rm 30\,mas}$ could originate from the 
very nucleus of NGC 1068 rather than from the torus. 
This could be achieved in a scattering halo above
and below the nuclear torus. In this halo a large part of the flux could be
isotropically scattered rather than absorbed and thermalized in an opaque
torus along our direct line-of-sight to the nucleus. $F_{\rm K}^{\rm 30\,mas}$ 
lies only about a factor 
of two above the extrapolated $F_\nu \propto \nu^{1/3}$ spectrum derived 
for the range around 10\,GHz: The spectral index ${\rm d}F_\nu / {\rm d}\nu$ 
between 5\,GHz and the K band amounts to $\le 0.39$. We note that this value 
is very similar to that of other galactic nuclei, like Sgr A* (BDM96), M\,81 
(Reuter and Lesch 1996); M\,104 (Jauch and Duschl, in prep.), 
where $\alpha = 1/3$. If NGC 1068 has the same spectral shape as these other 
galactic nuclei, then a fraction of $F_{\rm K}^{\rm 30\,mas}$ could indeed be 
contributed from the nucleus of NGC 1068. 

However, one has to admit that very little is known about the true nuclear 
spectrum of NGC\,1068 in the intermediate frequency range. To persue our 
speculation, we assume -- as a working hypothesis -- that also between 22\,GHz 
and the IR range, the spectrum goes like $\nu^{1/3}$. We then follow BDM96 and 
interpret this as optically thin synchrotron radiation of 
quasi-monoenergetic electrons.
The mean electron energy then is fairly well constrained since the maximum of 
$F_\nu^{\rm nuc}$ has to be 
at frequencies above the K band, but not much higher as otherwise the total 
nuclear flux from the center of NGC\,1068 would be too large. The situation is 
less clear with the SSA frequency. We cannot 
rule out that SSA in fact occurs at frequencies even smaller than 5\,GHz. As a 
consequence of this, the source radius discussed below is only a lower limit. 
For details we refer the reader to BDM96\footnote{For other 
models of such a spectral distribution, we refer to the discussion in MDZ96}.

\begin{figure}
\centerline{\resizebox{0.85\hsize}{!}{\includegraphics{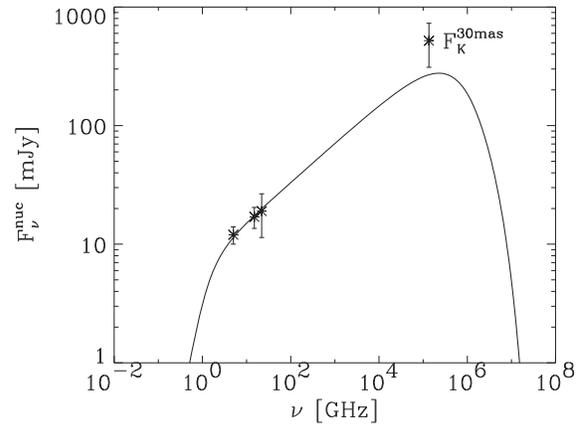}}}
%\vspace{-3mm}
\caption{A comparison of the model spectrum (for parameters, see the text)
with flux determinations with a sufficiently high spatial resolution. The
fluxes were taken from MPH96 (5 and 22\,GHz), Ulvestad, Neff, Wilson \ 
(1987; 15\,GHz), and the present paper (K-band). \label{figspect}}
\end{figure}

If we assume that the maximum of $F_\nu^{\rm nuc}$ is indeed achieved around 
2\,$\mu$m, and that SSA of the source becomes important for frequencies below 
5\,GHz, we find as emitting region a homogeneous sphere of radius $R \sim 
2\,10^{15}\,$cm ($\sim 0.7\,{\rm 
mpc}\sim 130\,{\rm AU\sim 0.01\,mas}$) with a magnetic field $B \sim 11\,$G 
(assumed to be the same everywhere in this region). The relativistic electrons 
have a number density $n_{\rm e} \sim 1.1\,10^3\,{\rm cm}^{-3}$, a mean 
energy $E \sim 2.7\,$GeV and a width of the energy distribution $\Delta E / 
E \sim 1$. In Fig.\ \ref{figspect}\ we show a comparison of the observed 
fluxes of NGC\,1068 core and our model spectrum using the above parameters.
If our speculation applies, it 
turns out that the main difference between NGC\,1068 and other galactic 
centers analysed on the basis of the same interpretation (Sgr A*: BDM96; 
M\,81: Reuter and Lesch 1996; M\,104: Jauch and Duschl, in prep.) 
are the source radius and - especially - the energy of the relativistic 
electrons. 
The above size of 0.01\,mas means that our resolved 30\,mas object is not
the synchrotron source itself but rather a larger object, most likely the 
nuclear torus and/or a circumnuclear scattering halo.

\section{Conclusions}
We have resolved a compact source with a diameter of $d_{\rm Gauss}\sim 
30\,$mas in the core of
NGC\,1068. This object is most likely a nuclear torus and/or a
circumnuclear scattering halo. Part of the radiation from the central 30\,mas may
be light from the nucleus scattered in the halo. Under this assumption, we 
were able to determine physical parameters of the nucleus and compare them to other
galactic centers, active and non-active ones.
One then is tempted to speculate that the higher efficiency of the 
acceleration mechanism for the electrons may be the true difference between an 
active galactic center and a normal one.

\begin{acknowledgements}
We thank Ms.\ Si$\hat{\rm a}$n Adey for valuable comments on the manuscript and
the referee for many helpful comments concerning the flux determination and 
other comments.
This work was in part supported by the {\it Deutsche Forschungsgemeinschaft\/} 
through Sonderforschungsbereich 328 ({\it Evolution of galaxies\/}) at the 
University of Heidelberg.
\end{acknowledgements}

\end{document}